\documentclass[aps,prb,preprint,superscriptaddress]{revtex4}
\usepackage{graphics}
\usepackage{graphicx}
\usepackage{amsmath}
\usepackage{amssymb}
\usepackage{bm}
\usepackage{dcolumn}

\begin{document}  

   \title[]{Electronic and optical properties of InAs(110)}   
   \author{X. L\'opez-Lozano} 
   \affiliation{Instituto de F\'{\i}sica, Universidad Aut\'onoma de Puebla, Apartado Postal J-48, Puebla 72570,  M\'exico}  

   \author{Cecilia Noguez}
   \email[Author to whom correspondence should be addressed. Email:]{cecilia@fisica.unam.mx} 
   \affiliation{Instituto de F\'{\i}sica, Universidad Nacional Aut\'onoma de M\'exico, Apartado Postal   20-364, Distrito 
   Federal 01000,  M\'exico}  

   \author{L. Meza-Montes} 
   \affiliation{Instituto de F\'{\i}sica, Universidad Aut\'onoma de Puebla, Apartado Postal J-48, Puebla 72570,  M\'exico}  
	  
   \date{\today}  

 \begin{abstract}
The electronic and optical properties of the cleavage InAs(110) surface are studied using a semi-empirical tight-binding method which employs an extended atomic-like basis set.  We describe and discuss the electronic character of the surface electronic states and we compare with other theoretical approaches, and with experimental observations. We  calculate the surface electronic band structure and the Reflectance Anisotropy Spectrum, which are described and discussed in terms of the surface electronic states and the atomic structure. 
 \end{abstract}

 \maketitle 

\section{Introduction}
\label{}

The (110) surface is the natural cleavage of zincblende crystals, and it is a non-polar surface which contains equal number of cations and anions in its unit cell, showing a partly ionic bonding. The mechanism of reconstruction and the electronic properties of the (110) III-V semiconductors seems to be broadly understood~\cite{opto,GaAs,godin} but this is not in detail, and in fact, they have different optical properties~\cite{pulci1,pulci2,noguez2,cappe}. A lot of theoretical work has been done to determine the electronic properties of Gallium compounds, and their surfaces. On the other hand, only  few attempts have been made to characterize the InAs surfaces. In this work, we are interested in the InAs(110) surface, and in the manuscript, we show that there is not agreement between previous theoretical results, and between those theoretical results and experimental measurements reported in the literature of the InAs(110) surface. Therefore, more theoretical and experimental studies are necessary. 

Most of the theoretical~\cite{alves,Mail,beres,engels,anatoli} studies about InAs(110) do not provide a way for directly compare with experiments~\cite{and1,and2,SwanO,Swan,leed,leed1,richter,carstensen,Drube,gudat,laar}. Furthermore, the available experimental measurements are not enough to completely elucidate the atomic structure and electronic properties of InAs(110).  For example,  Andersson and collaborators~\cite{and1} found  the energies of occupied-surface states at high-symmetry points using photoemission techniques. Six years later, the same group measured again the occupied-surface states at the same high-symmetry points~\cite{and2}, founding  a systematic shift of energy of $-0.15$~eV with their previous measurements~\cite{and1}. Independently, Swantson {\it et al.}~\cite{Swan} also measured the energies of occupied-surface states at high-symmetry points and they found differences up to 0.5~eV with those reported by Andersson and collaborators~\cite{and2}. In general, the interpretation of photoemission spectra has been difficult to do because of the small bulk-band gap of InAs.

Theoretically, the electronic structure and atomic positions of InAs(110) were calculated using a quantum-molecular dynamics~\cite{Mail} based in a semi-empirical tight-binding (TB) approach. Almost a decade latter, an {\em ab initio} quantum-molecular dynamics~\cite{alves} was performed. The reported atomic structure and electronic surface states differ between TB  and {\em ab initio} calculations, and also differ with the available experimental measurements. This fact is because the semi-empirical calculations were performed using an atomic reconstruction that was not fully relaxed. The {\em ab initio} calculation was performed using DFT-LDA with a plane-wave basis set whose accuracy was compromised with the choice of several approximations like the energy cut-off, obsolete pseudopotentials, etc. Therefore, the {\em ab initio} calculation~\cite{alves} presented systematic errors to determine the surface electronic energy levels. Although {\it ab initio} methods are better than semi-empirical ones,  it is well-known that such methods are far to be easily implemented.  For example, it is known that for small band gap semiconductors crystals, like Ge, InAs, etc., {\it ab initio} methods predict a metallic behavior, because the gap results to be negative. It is also known that ab-initio methods need to use a large cut-off energy  to achieve convergence in surface electronic states, and incorporate many-body electron interactions to get the right dispersion of them.  Then, it is expected that semi-empirical methods are more suitable than ab initio calculations for the InAs(110) surface. We believe that for this reason only one {\it ab initio} calculation is found in the literature and it is twelve years old.

The optical properties of InAs(110) have been also investigated both, theoretically and experimentally. Shkrebtii and collaborators~\cite{anatoli} calculated and measured the Reflectance Anisotropy Spectrum (RAS) of InAs(110). Although the authors claimed to elucidate the optical properties of such surface, the calculated RAS is far to resemble their measurements. The discrepancies between their theoretical results and measurements can be associated again to the small optical bulk gap of InAs, which ``hides'' the main features of the optical spectrum coming from electronic transitions which involve surface states.  In summary, with the available theoretical and experimental evidence it is no possible to clearly elucidate the main electronic and optical properties of  InAs(110).  In this work, we study the electronic structure and optical properties of InAs(110) employing a semi-empirical TB  formalism~\cite{noguez,Vogl} and using the atomic coordinates obtained from {\em ab initio} quantum-molecular dynamics~\cite{alves}. The use of the fully relaxed atomic coordinates guarantees that the calculated electronic properties include all the subtle effects of surface-induced strain and appropriate geometry. The TB approach allows us to  analyze in detail the electronic structure and optical properties, and compare our calculations with available experimental data. 

\section{Theoretical method}

The III-V(110) semiconductor surfaces relax in such a way that the surface cation atom moves inwards the surface into an approximately planar configuration, with a threefold coordination with its first-neighbors anion atoms. The topmost anion atom moves outward to the surface, showing a pyramidal configuration with its three first-neighbors cation atoms~\cite{GaAs,godin}.  The geometric parameters that describe the relaxation of the surface atoms of III-V(110) semiconductor surfaces, scale linearly with the bulk lattice constant~\cite{godin}. In particular, Alves {\it et al}.  found~\cite{alves} that for the InAs(110) surface the {\it pyramidal} angle at the anion, labeled by $\alpha$,  is $\sim  90^{\circ}$, the {\it in-plane} angle $\beta$ has values close to the tetrahedral bond angle $\sim 109.47^{\circ}$, and the {\it planar} angle at the cation, labeled by $\gamma$, is $ \sim 120^{\circ}$. For the ideal surface the values for $\alpha$, $\beta$ and $\gamma$ correspond to those angles of tetrahedral bonds, $109.47^{\circ}$.


The relaxed InAs(110) surface is shown in Fig.~1. In Fig.~1(a) we show the top view of a surface unit cell that contains one In atom (cation), and one As atom (anion) per atomic layer. The open circles correspond to As atoms while  black circles show In atoms. The parameter $a_{0}$ is the bulk lattice constant and $d_{0}= a_{0}/2\sqrt{2}$. The larger side of the unit cell is along the [001] crystallographic direction, while the shorter side is along the [$1\overline{1}0$]. In Fig.~1(b) we show a side view with only the three outermost atomic layers of the surface. Here, we define the structural parameters associated to the surface relaxation whose values are given in Table~1. In Fig.~1(c) we show  the corresponding Two-Dimensional Irreducible Brillouin Zone (2DIBZ).  


In our calculations, the non-polar InAs (110) surface was modeled using a slab of 50 atoms, yielding a free reconstructed surface on each face of the slab. The thickness of the slab is large enough to decouple the surface states at the top and bottom surfaces of the slab. Periodic boundary conditions were employed parallel to the surface of the slab to effectively  model an infinite two-dimensional crystal system. The atomic coordinates were taken from Ref.~(8), and are given in Table~1. We have performed calculations with all the structural parameters in Table~1, however, those corresponding to the Density Functional Theory~\cite{alves} (DFT) with an energy cutoff of 18~Ry are the ones that best resemble some  experimental data~\cite{and1,and2,SwanO,Swan,leed,leed1,richter,carstensen,Drube,gudat,laar}. We calculate the electronic level structure of the slab using a well known parameterized TB approach with a sp$^3$s$^*$ orbital-like basis, within a first-neighbor interaction approach~\cite{Vogl}. This wave function basis provides a good description of the valence and conduction bands of cubic semiconductors, except in the X -- W direction in the Brillouin zone where the conduction bands are underestimated. This TB approximation has been applied to calculate the electronic and optical properties of a variety of semiconductor surfaces, including other III-V compounds~\cite{noguez2}. The TB parameters are taken to be the same as those of Vogl~\cite{Vogl} for the bulk but they are scaled by a factor of $(D/d)^2$, where $d$ is the bond length of any two first-neighbor atoms, and $D = \sqrt{3}a_0/4$.~\cite{harrison} These changes to the original bulk parameters provide an excellent description of the electronic structure, as compared to experimental measurements.  

Once the electronic-level structure of the slab has been obtained, we calculate the average slab polarizability, which is in terms of the transition probabilities between eigenstates induced by an external radiation field. We take an average over 7000 points distributed homogeneously in the irreducible two-dimensional Brillouin zone (2DBZ). The real part of the average polarizability is calculated using the Kramers-Kronig relations. Finally, the RAS is calculated as the difference of the Differential Reflectance between two orthogonal directions in the surface plane, as
\begin{equation}
{\rm RAS} = \left(\frac{\Delta R }{R_0}\right)_{[1\overline{1}0]} - \left(\frac{\Delta R }{R_0} \right)_{[001]},
\end{equation}
where $R_0$ is the bulk reflectivity calculated with the well-known Fresnel formula, and $\Delta R = R - R_0$ is the difference between $R_0$ and the actual reflection coefficient. The details are fully explained in Ref.~(24).  

The atomic structure of the surface region is intimately related to its electronic structure. Experimentally, the electronic structure can be determined by means of electron spectroscopies like photoemission (PE), inverse photoemission (IPE), and Scanning Tunneling Spectroscopy (STS). These techniques are sensitive to the surface's features and electronic properties due to reconstructions or adsorption events. We present and discuss in Section~III the surface electronic band structure and the local density of electronic states of InAs(110), and in Section IV, we discuss the results of the optical properties. 

\section{Surface Electronic Structure}
\subsection{Results}
We show the surface electronic band structure along high-symmetry points of the 2DIBZ of the reconstructed InAs(110) surface in  Fig.~2. The projected bulk electronic states are shown in tiny black dots, while the surface electronic states are shown in large black dots. We denote the surface electronic states  using the labels $A_{i}$ and $C_{i}$ associated to the surface anions and cations, respectively, as introduced by Chelikowsky and Cohen~\cite{chel}. The calculated Fermi energy level, ${\rm E_F}$, is at 1.1~eV above the Valence-Band-Maximum (VBM), while  the Minimum of the Conduction Band (MBC)  is at $0.6$~eV from the VBM at the high-symmetry point $\Gamma$.  Notice that the measured ${\rm E_F}$ of bulk InAs is found to be at about 0.55~eV, and the measured minimum bulk gap is about 0.35~eV. For the bulk InAs,  we obtained that the bulk ${\rm E_F}$ is about 0.6 eV, in agreement with experimental observations~\cite{and2}. For InAs(110) there is not agreement between experimental measurements of ${\rm E_F}$, however, all of them show that its value increases respect with the bulk one. Therefore, it is not surprising that ${\rm E_F}$ seems to be in the conduction band.  


Below the  VBM we found four well-defined occupied surface electronic states denoted by  $A_{5}$, $A_{3}$, $A_{2}$ and $C_{2}$.  The $A_{5}$ surface states correspond to the dangling bonds of the As atoms located at the first atomic layer. The $A_5$ states form a band from the high-symmetry point X to the point ${\rm X}'$, going through the high-symmetry point M in the 2DIBZ. This band has a minimum at X with an energy of -1.20~eV and disperses upwards towards the $\Gamma$ point. From X, the band also disperses upwards towards the M point, where the $A_{5}$ surface states have an energy of about -0.8~eV.  From M to X, this band disperses into the projected bulk band. The $A_5$ band shows a small dispersion around M given rise to a large contribution on the Local Density of States (LDOS) of the first layer at an energy of about -1~eV, as shown in Fig.~3.  


The $A_{3}$ surface electronic states are at a lower energy than $A_5$. The $A_3$ states are due to the backbonds between the anions localized in the first atomic layer and the cations in the second layer. The $A_3$ band has a minimum in the X high-symmetry point with an energy of -2.8~eV from the VBM. The band reaches its maximum at ${\rm X}'$ with an energy of -1.6~eV from  the VBM. The band shows a dispersion of 1.2~eV, however, around X and M the band is almost flat, contributing to a large density of states in the first and the second layers at energies of about -2.8~eV and -2.5~eV, respectively. This can be observed on the  LDOS  in Fig.~3, where two peaks are found at these energies in the panels showing the  LDOS in the first and the second layers. 

We found surface states with an energy of about $-6.0$~eV at the high-symmetry point X, that  form a band denoted by $C_2$. This  band shows a large dispersion of about  2.5~eV, where the minimum of the band is at M with an energy of -6.3~eV, and its maximum is around ${\rm X}'$ with an energy of -3.8~eV. These surface states are located at the cation (In) atoms and are due to the bonding between the In and As atoms at the top layer. From X to M, the band shows a small dispersion which is reflected in the LDOS where a large contribution is found at about -6.2~eV in the panel showing the projected  LDOS in the first layer in Fig.~3. From M to ${\rm X}'$, the $C_2$ band disperses upwards of about 2.5~eV given rise to a small contribution to the LDOS as shown in Fig.~3. 

At lower energies we found another occupied surface electronic band denoted  by $A_2$ which extends almost along all the high-symmetry points in the 2DIBZ. These surface states are located in the anion (As) atoms, and have a $s$ character due to  the backbonds between the atoms at the first and second layers, and some contribution is also found from the backbonds between the atoms in the second and third layers. From $\Gamma$ to M going along X, the band does not show dispersion and is at $-10.2$~eV. From M to $\Gamma$ going along ${\rm X}'$, the band has a dispersion of about 1~eV, showing a minimum around ${\rm X}'$. 
  
We have also found several localized states at $X$ between $\Gamma$ and $M$ with an energy from -2~eV to -1~eV. These states are inside the projected bulk band, therefore, they are resonance-like states. These resonance states, denoted by $A_4$, disperse upwards from $X$ towards both, $\Gamma$ and $M$. Most of the $A_4$ states have a $p$-character, and are localized at the anion in the first atomic layer. The $A_4$ states with lower energy, between  -1.9~eV and -2.0~eV,  show also a $s$ character and are localized at the third layer.  

Above the VBM we found two unoccupied surface states bands, namely, $C_3$ and $C_4$. The $C_3$ surface states are localized at the cations in the first and third  atomic layers. They show a strong $p$ character due to the dangling bonds at cations. At X, we found that $C_3$ has a maximum with an energy of about 2~eV, and has its minimum value between M and ${\rm X}'$ with a energy of about 1.4~eV.  Finally, at 2.7~eV from VBM we found empty surface states that form a band along all the high symmetry points in the 2DIBZ.  This band is denoted by $C_4$, and show a very small dispersion along the 2DIBZ.

\subsection{Discussion}

In this section we discuss our results, and compare with available experimental measurements~\cite{and1,and2,SwanO,Swan,leed,leed1,richter,carstensen,Drube,gudat,laar} and theoretical calculations~\cite{alves,Mail,beres,engels,anatoli}. The electronic properties of InAs(110) have been investigated previously using experimental techniques like photoemission (PE)~\cite{and1,and2,SwanO,Swan,richter,carstensen} and inverse photoemission (IPE)~\cite{carstensen,Drube,gudat,laar} spectroscopies. We found no agreement between experimental measurements because they present difficulties to identify the position of the  VBM or ${\rm E_F}$, and the employed samples are quite different. Furthermore, photoemission measurements are difficult to interpret since emissions from surface states are usually hidden by emissions from bulk states in InAs surfaces, due to the small bulk-band gap. 

On the other hand, theoretical calculations have been performed from {\it ab initio}~\cite{alves} and semi-empirical TB~\cite{Mail,beres,engels,anatoli} methods. We summarize our results and some of the experimental and theoretical data in Table~2, where we show the energy values at high symmetry points in 2DIBZ of the surface electronic states denoted by $A_{5}$, $A_{3}$, $A_{4}$ and $C_{2}$. The first column shows our results, the next three columns show experimental measurements obtained by PE~\cite{and1,and2,Swan}, and the last two columns show theoretical results~\cite{alves,Mail}. The values in Table~2 are those reported in the corresponding reference, or they have been estimated from the figures in each reference, then errors of about 0.1~eV in the estimated values are expected.  


Alves {\it et al.}~\cite{alves} performed an {\it ab initio} calculation based in the Density Functional Theory (DFT) within the Local Density Approximation (LDA), where many-body effects were not taken into account. They considered slabs of only eight atomic layers (16 atoms), and the employed plane-wave basis set was expanded up to an energy cutoff of 8 and 18~Ry. They reported results for the equilibrium atomic structure and the electronic band structure. It is known that the equilibrium atomic geometries can be found with good accuracy, but underestimation and/or overestimation of  electronic states  is always present in DFT-LDA calculations due to the approximations employed, for example, usually DFT neglects many-body effects~\cite{mb1}. Furthermore, plane-wave basis expansions always present convergence problems to find localized states. Therefore, the comparison of the surface states from {\it ab initio} calculations ~\cite{alves} with semi-empirical results and PE measurements always presents deviations up to $\pm$1~eV. We also compare our results with semi-empirical TB calculations done by Mailhiot {\it et al.}~\cite{Mail}. These TB calculations employed a theoretical method similar to the one used here but with different atomic positions, that were not fully relaxed. 

Both theoretical calculations~\cite{alves,Mail} found  an empty surface state $C_3$ and was identified with dangling bond states at cations. While the DFT calculation~\cite{alves} found that $C_3$ has a minimum at X, we obtained a maximum at the same symmetry point in agreement with other semi-empirical calculations~\cite{Mail,beres,engels,anatoli}. DFT calculations reported an upwards dispersion from X of about 1.4~eV, while we found a downwards dispersion from X of about 0.6~eV. This discrepancy between DFT and semiempirical calculations is expected since DFT uses a small basis set that  can not reproduce conduction states, while semiempirical calculations with an extended basis set can do. The $C_{3}$ surface states have been measured by using inverse photoemission~\cite{carstensen,Drube,gudat,laar} but only at the X high-symmetry point. The experimental measurements assigned an energy between 1.7~eV and 1.9~eV at X, and evidence of an upwards dispersion from this point have been observed~\cite{carstensen} in agreement with our calculations. However, a more detailed experimental analysis is necessary to conclude more about empty states. 

The occupied surface states denoted by $A_5$, $A_3$, and $C_2$, were also calculated in Refs.~(8) and (9). Both calculations identified the $A_5$ surface states with the dangling bond states in the anion atoms, in agreement with our results. The $A_5$ surface state found using a first-principles method~\cite{alves} is shift 0.3~eV in average, above the experimental value~\cite{and2} and our calculation. Experimentally~\cite{and2}, the $A_5$ surface state has a dispersion of about 0.15~eV from X to ${\rm X}'$ through M, while we calculate a dispersion of 0.5~eV, similar to the one obtained using DFT~\cite{alves}. The surface states denoted by $A_4$ have been also observed experimentally~\cite{and1,and2}, and calculated by Mailhiot {\it et al}.~\cite{Mail}. Like for the $A_5$ states, we can not identify $A_4$ states at the $\Gamma$ point. PE measurements showed that these states have a dispersion of about 0.55~eV from X to to ${\rm X}'$ through M, which is in agreement with our calculated value of 0.5~eV, while Mailhiot {\it et al}.~\cite{Mail} found a smaller dispersion of 0.2~eV, and this value was not calculated using DFT~\cite{alves}. Previous semiempirical results~\cite{Mail} also found the occupied states labeled by $A_4$, however, these surface states were not well identified since in their calculations the $A_4$ states show a very similar dispersion than the $A_5$ states.   In general, the $A_4$ are resonant states and they are difficult to calculate, specially if a plane-wave basis is used as the DFT calculations discussed here~\cite{alves}. Below $A_4$, it has been observed other resonant states denoted by $A_3$, using PE~\cite{and2}. The reported data for $A_3$ is quite different from our calculations and previous TB calculations~\cite{Mail}, maybe because the identification of these states is not clear experimentally~\cite{and1,and2}. On the other hand, the $C_2$ states can be experimentally identified since they are in a gap, except at the X point where they disperse into the projected bulk states. 

\section{Optical Properties: 
Results and Discussion}

We caculate RAS according to Eq.~(1) for InAs(110) as a function of the energy of the incident light. RAS has contributions of electronic transitions from occupied to empty states which are labeled as surface to surface ({\em ss}), surface to bulk ({\em sb}), bulk to bulk ({\em bb}), and bulk to surface ({\em bs}) electron transitions. The bulk-band gap is about 0.5~eV, such that {\em bb} transitions hide most of the electron transitions involving surface states. Therefore, we only consider {\em ss, sb}, and {\em bs} transitions to calculate RAS, as shown in Fig.~4.  In Fig.~4(a) we show our calculated RAS (solid line), the measured (dotted line) and calculated (dashed line) RAS in Ref.~(12). The measured curved was found to be positive from 1.5 to 4.5~eV, however, we plotted it with a shift of $-0.7\%$ to easily compare the experimental measurements with our results and the theoretical results from Ref.~(12). We believe that this difference comes from the fact that we are not considering {\em bb} transitions. In Fig.~4(b), we show our calculated RAS denoted by total and its electron-transition contributions. In general, our calculated RAS resembles some of the main features of the measured RAS more than a previous calculated spectrum from Ref.~(12), also shown in Fig.~4(a).


We found a minimum in the spectrum for energies between 1.5 and 2.2~eV, in agreement with measured RAS~\cite{anatoli}. We observe that {\em sb} and {\em bs} electron transitions  below 2.2~eV are more intense along the [001] direction than their contributions along [1$\overline{1}$0]. A broad structure around 2.5~eV and about 0.75~eV wide is observed in our calculated RAS.  In Ref.~(12) they related this peak to {\em ss} transitions, while we found that such peak has contributions from all kind of electron transitions involving surface states, being {\em sb} the dominant ones, although some {\em bs} and {\em ss} transitions are also present. The {\em sb} transitions are related with the $A_5$ surface states due to the dangling bonds in the As atoms, and bulk empty states, and the electron transitions occur mainly along the M to X'/Y high-symmetry points which show such dispersion. This wide structure is observed experimentally, where an additional shoulder appears at 2.75~eV which we assign to the {\em sb} transitions, while Shkrebtii {\em et al.}~\cite{anatoli} associated it to modifications of bulk states at the $E_1$ critical point. On the other hand, at the same energy (2.5~eV), RAS has also {\em bs} contributions coming mainly from the valence bulk states between the X and M high-symmetry points to $C_3$ empty surface states, where such surface states are due to the dangling bonds in the In atoms. Shkrebtii and collaborators~\cite{anatoli} had associated the peak at 2.5~eV to {\em ss} transitions, while we found that it is more related to {\em sb}. The {\em ss} transitions occur mainly at about 2.5, 3.2 and 3.4~eV. The first two peaks at lower energy are transitions from $A_5$, and $A_4$ surface states to $C_3$ surface states located at the dangling bonds at the surface As and In atoms, respectively. These peaks correspond to {\em ss} transitions along the surface chains in the [$1\overline{1}0$] direction, in such a way, they are positive, while the third peak at 3.4~eV is negative and is due to transitions between $A_3$ to $A_4$ surface states. We found a second positive peak at 3.2~eV which corresponds to the measured one at the same energy. Shkrebtii {\it et al}. found also this peak but at 3~eV and they found that the peak was mainly related to {\em bb} transitions. They also found~\cite{anatoli} that the main features of RAS at higher energy are due to {\em bb} transitions only. This conclusion is expected since in their calculations the {\em bb} transitions hide the contributions from surface states. For energies above 3.5~eV we cannot found correctly the optical features, since our atomic-like basis is not able to reproduce the E$_2$ bulk transition at about 4~eV. 

\section{Summary}

We performed a tight-binding calculation using a fully relaxed atomic geometry to study the electronic structure and optical properties of the clean InAs(110) surface. A very detailed analysis was done, and a good agreement between our calculations and some experimental data was found. In conclusion, we found that fully relaxed atomic positions from {\it ab initio} methods in combination with our semi-empirical tight-binding calculation, better resembles photoemission measurement and Reflectance Anisotropy Spectrum of cleavage InAs(110) samples.  We explain the main features of the optical spectrum between 2~eV to 4~eV, and relate them to the surface atomic and electronic structure.  Although the agreement between our results and other theoretical and experimental data is good, we conclude that more experimental studies are necessary to clearly elucidate the atomic relaxation and the electronic properties of InAs(110).

\acknowledgments We acknowledge the partial financial support from DGAPA-UNAM grant No.~IN104201, CONACyT grants No.~36651-E  and 36764-E.

\newpage

\begin{table} 
\begin{center} 
\vskip1cm
\begin{tabular}{|l|c|c|c|c|c|}
\hline \hline
 &$a_{0}$ (\AA)& $\Delta_{1,\perp}$ (\AA)& $\Delta_{1,x}$(\AA)&$\Delta_{2,\perp}$(\AA)& $d_{12,\perp}$(\AA) \\
\hline 
Ideal & $6.04$& $0.0$&$3/4a_{0}$ &$0.0$ &$d_{0}$  \\
DFT* & 5.844 & 0.70 & 4.656 & 0.122 & 1.463 \\
DFT**& 5.861 & 0.75 & 4.663 & 0.128 & 1.445  \\
LEED & 6.036 & 0.78 & 4.985 & 0.140 & 1.497 \\ 
\hline 
 & $d_{12,x}$(\AA)&$\omega$ (deg) & (\%)$c_{1}a_{1}$ &(\%) $c_{2}a_{1}$&(\%)$c_{1}a_{2}$\\
\hline
Ideal  & $a_{0}/2$&$0.0$ &$0.0$ &$0.0$ &$0.0$ \\
DFT*  & 3.361 & 30.7 & $-1.80$ & $-0.22$ & $-2.00$ \\
DFT**  & 3.395 & 32.0 & $-1.18$ & $-0.18$ & $-1.82$ \\
LEED & 3.597 & 36.5 & $-4.22$ & $+2.03$ & --- \\
\hline \hline
\end{tabular}
\end{center} 
\caption 
{Structural parameters as defined in Fig.~1. Parameters obtained from DFT calculations using an energy  cutoff of 8~Ry (*), and 18~Ry (**), both from Ref.~(8), and parameters from Low-Energy Electron Diffraction (LEED) measurements from Ref.~(22).} 
\label{parameters}
\end{table}

\begin{table} 
\begin{center} 
\begin{tabular}{|l|c|c|c|c|c|c|}
\hline \hline
State & This work & PE~\cite{and1}& PE~\cite{and2}&PE~\cite{Swan}&DFT~\cite{alves}&TB$^*$~\cite{Mail}\\
\hline 
$A_{5}(\Gamma)$ &  & $-0.30$ &$-0.45$ &$-0.53$ & &$-0.3$ \\
$A_{5}(X)$ & $-1.21$ & $-1.00$ & $-1.15$ & $-0.83$ &$-0.85$ & $-0.9$\\
$A_{5}(${\rm X}'$/Y)$ & $-0.70$ & $-0.85$ & $-1.00$ & $-0.73$ & & $-0.7$\\
$A_{5}(M)$ & $-0.81$ &  & $-1.10$ & $-0.70$ &  & $-0.8$\\
\hline
$A_{3}(\Gamma)$ &  &  & $-2.60$ &  & & $-2.1$\\
$A_{3}(X)$ & $-2.81$ & $-3.1$ & $-3.25$ & $-2.72$ &$-3.21$ & $-3.1$\\
$A_{3}(${\rm X}'$/Y)$ & $-1.57$ &  & $-1.50$ &  & &$-1.4$ \\
$A_{3}(M)$ & $-2.51$ &  & $-3.40$ &  & & $-2.7$\\
\hline
$C_{2}(\Gamma)$ &  &$-3.35$ & $-3.50$ &  & & $-3.2$\\
$C_{2}(X)$ & $-6.04$ &  & & $-4.90$ & & $-5.8$ \\
$C_{2}(${\rm X}'$/Y)$ & $-3.8$ &$-3.7$  & $-3.85$ & $-3.35$ &$-3.4 \dagger$ &$-3.6$ \\
$C_{2}(M)$ & $-6.28$ & & $-6.10$ &  &$-5.46$ &$-6.1$ \\
\hline 
$A_{4}(\Gamma)$ & & &$-0.45$ & & & $-0.6$\\
$A_{4}(X)$ & $-1.53$ & $-1.6$ & $-1.75$ & & & $-1.2$\\
$A_{4}(${\rm X}'$/Y)$ & $-1.36$ & $-1.05$ & $-1.20$ & & & $-0.9$\\
$A_{4}(M)$ &$-1.03$ &  & $-1.10$ & & & $-1.0$\\
\hline \hline
\end{tabular}
\end{center} 
\caption{Experimental and theoretical values of the surface states at high-symmetry points of the 2DIBZ. The energy values are in eV, where the zero energy corresponds to the  VBM. ($\dagger$) Estimated value from Fig.~7(a) in Ref.~(8). (*)~Estimated values from Fig.~9(b) in Ref.~(9). An error of about 0.1~eV is expected in the estimated values.} 
\label{results}
\end{table}

\begin{figure} [t]
\begin{center}
\end{center}
\caption{Model of the atomic geometry of InAs(110). (a)~Top view of a surface unit cell. (b)~Side view of the first three atomic layers of the surface. (c)~Two-Dimensional Irreducible Brillouin Zone.}
\label{atomos}
\end{figure}

\begin{figure}[t]
\begin{center}
\vskip1in
\end{center}
\caption{Electronic band structure of the reconstructed InAs(110) surface. Tiny dots represent the projected bulk states, while black dots represent surface electronic states.}
\label{ebclean}
\end{figure}

\begin{figure}[htb]
\begin{center}
\end{center}
\caption{Total and the projected local density of states in the first, second and third atomic layers of reconstructed InAs(110).}
\label{dosclean}
\end{figure}

\begin{figure}[htb]
\begin{center}
\end{center}
\caption{(a) Reflectance Anisotropy Spectrum calculate by us (solid line), and measured (dotted line) and calculated (dashed line) in Ref.~(12).  (b) Our total calculated RAS (solid line) and its {\em ss} (dotted line), {\em sb} (dashed line), and {\em bs} (dotted-dotted-dashed line) components.} \label{ras}
\end{figure}

\end{document}